\documentclass[aps,prl,reprint,amsmath,amssymb,superscriptaddress,floatfix]{revtex4-1}
\usepackage{graphicx}
\usepackage[colorlinks, citecolor=blue, linkcolor=blue]{hyperref}

\begin{document}

\title{Entropy-reduced retention times in magnetic memory elements: A case of the Meyer-Neldel Compensation Rule}

\author{L. Desplat}
\email{louise.desplat@ipcms.unistra.fr}
\affiliation{Centre de Nanosciences et de Nanotechnologies, CNRS, Universit{\'e} Paris-Saclay, 91120 Palaiseau, France}
\affiliation{Institut de Physique et Chimie des Mat{\'e}riaux de Strasbourg, CNRS, Universit{\'e} de Strasbourg, 67200 Strasbourg, France}

\author{J.-V. Kim}
\affiliation{Centre de Nanosciences et de Nanotechnologies, CNRS, Universit{\'e} Paris-Saclay, 91120 Palaiseau, France}
\date{\today}

\begin{abstract}
We compute mean waiting times between thermally-activated magnetization reversals in a nanodisk with parameters similar to a free CoFeB layer used in magnetic random access memories. By combining Langer's theory and forward flux sampling simulations, we show that the Arrhenius prefactor can take values up to 10$^{21}$ Hz, orders of magnitude beyond the value of 10$^{9}$ Hz typically assumed, and varies drastically as a function of material parameters. We show that the prefactor behaves like an exponential of the activation energy, which highlights a case of the Meyer-Neldel compensation rule. This suggests that modeling information retention times with a barrier-independent prefactor in such magnetic storage elements is not justified.
\end{abstract}

\maketitle


The Meyer-Neldel (MN) rule, also known as entropy-enthalpy compensation or the isokinetic rule, has been known empirically to chemists since the 1920s~\cite{constable1925mechanism} and to physicists since 1937~\cite{meyer1937relation}. It describes many processes across the natural sciences~\cite{yelon2006multi}, such as biological death rates~\cite{rosenberg1971quantitative}, transport in semiconductors \cite{meyer1937relation, kamiya2010present}, chemical reaction rates \cite{liu2001isokinetic}, and geological diffusion profiles~\cite{hart1981diffusion}. 
For thermally activated processes whose rate $k$ follows an Arrhenius-type law~\cite{hanggi1990reaction},
\begin{equation}
k = f_0 e^{-\beta \Delta E},
\label{eq:arrhenius}
\end{equation}
in which $\Delta E$ is the internal energy barrier, $\beta=(k_B T)^{-1}$ is Boltzmann's factor, and $f_0$ is the Arrhenius prefactor, then if the Meyer-Neldel rule applies  \cite{yelon1990microscopic,yelon1992origin,yelon2006multi}, 
\begin{equation}\label{eq:mn}
\ln f_0 = \ln f_{00} + \frac{\Delta E}{E_{0}} + b,
\end{equation}
where $E_{0}$ is a characteristic energy of the family of transitions, $f_{00}$ is a positive factor and $b$ is a constant. A general result is that processes described by Eq.~(\ref{eq:mn}) should possess a large activation energy compared to the thermal energy and the fundamental excitations in the system. The quanta of excitation of the heat bath are typically bosons, such as phonons in solid state physics, photons in the case of electronic transitions, or magnons in magnetism.  Since $\Delta F = \Delta E - T \Delta S$, where $\Delta F$ is the Helmholtz free energy, Eq.~(\ref{eq:arrhenius}) may be reformulated as
\begin{equation}
k=f_{00} e^{-\beta \Delta F}  = f_{00} e^{\Delta S/k_B} e^{-\beta \Delta E},
\end{equation}
and compensation follows if $\Delta S/k_B= \Delta E/E_0 + b$. This type of relation seems to  naturally arise in the case of large activation energies, by counting the number of ways the heat bath can furnish the energy needed to overcome the barrier
 \cite{peacock1982compensation}. Refs.~\onlinecite{yelon1990microscopic, yelon1992origin} find that compensation should be associated to multi-excitation processes.

Within the magnetism community, it is common practice to consider the attempt frequency as a characteristic timescale of the dynamics, i.e., $f_0 \sim 1$ GHz. The knowledge of the internal energy barrier then naturally leads to a direct estimation of the rate of thermally activated magnetic transitions via Eq.~(\ref{eq:arrhenius})~\cite{weller1999thermal, chen2010advances, lederman1994measurement, cortes2017thermal}. In particular, this is the case for the mean waiting time between magnetization reversals in nanostructures, such as nm-thick disks used in magnetoresistive random access memories (MRAM). Their stability can be evaluated as $\Delta = \beta_{300} \Delta E$, where $\beta_{300}$ is Boltzmann's factor at $T = 300$ K~\cite{khvalkovskiy2013basic}. A typical metric for information storage is a 10-year retention time, which requires a minimum of $\Delta \approx 50$. Recently, efforts to estimate the thermal stability of magnetic skyrmions have hinted that such an approach does not hold.  In systems with a multidimensional phase space, the Arrhenius prefactor cannot be interpreted as a literal ``attempt frequency'', because it also carries an important activation entropy~\cite{wild2017entropy, bessarab2018annihilation, desplat2018thermal, von2019skyrmion, desplat2020path}. Nevertheless, this effect has often been attributed to the nontrivial topology of magnetic skyrmions, rather than a general result~\cite{wild2017entropy, von2019skyrmion}. Meanwhile, more recent developments in MRAM have focused on storage elements with perpendicular magnetic anisotropy (PMA). The introduction, in the free layer, of elements with a strong spin orbit coupling, in order to enhance the PMA, has been shown to also induce a large interfacial  Dzyaloshinkii-Moriya interaction (DMI)~\cite{heinze2011spontaneous}. The typical configuration at the saddle point (SP) for the switching of the magnetization in such systems is two oppositely magnetized domains separated by a domain wall~\cite{sampaio2016disruptive}. The DMI selects a preferred chirality of the wall and lowers its energy, thus leading to lower activation energies and, from the assumption $f_0 \sim  1$ GHz, dramatically reduced retention times~\cite{sampaio2016disruptive, gastaldo2019impact}.

In this Letter, we show by applying Langer's theory~\cite{langer} and forward flux sampling (FFS)~\cite{allen2004sampling, allen2006simulating} that the prefactor for magnetization reversals in nm-thick disks can take extreme, seemingly non-physical values and vary drastically as a function of material parameters. We find that it behaves like an exponential of the activation energy, which stems from a linear variation of the activation entropy with the energy barrier, and thereby demonstrates a case of the Meyer-Neldel compensation rule.


To illustrate these compensation effects for magnetic memory elements, we follow the example in Ref.~\onlinecite{sampaio2016disruptive} and study a perpendicularly-magnetized CoFeB ultrathin film in a nanopillar within the micromagnetic approximation. The geometry comprises a disk of 32 nm in diameter with a thickness of $d = 1$ nm, which we model using finite difference cells $1 \times 1 \times 1$ nm$^3$ in size. We take an exchange constant of $A = 10$ pJ/m, a saturation magnetization of $M_s=1.03$ MA/m, and a variable interfacial DMI constant $D>0$. Dipolar interactions are treated in the local approximation through the use of an effective perpendicular anisotropy, $K = K_u - (N_z-N_x)\mu_0 M_s^2/2$, where $N_i$ are the demagnetizing factors of the disk~\cite{chen1991demagnetizing} and $K_u = 0.77$ MJ/m$^3$ is the base value. Below a critical DMI strength of $D_c = 4\pi^{-1}\sqrt{A K}$, the ground state is degenerate between uniformly-magnetized ferromagnetic (FM) states along the $+z$ (``up'') and $-z$ (``down'') directions. Above $D_c$, noncollinear states are favored, and the domain wall energy, $\sigma_w = 4 \sqrt{A K}-\pi D$, becomes lower than that of the FM state. Our goal here is to directly calculate the information retention time of the disk -- i.e., the mean waiting time, $\tau=k^{-1}$, between magnetization reversals.


\begin{figure}
\centering\includegraphics[width=8.5cm]{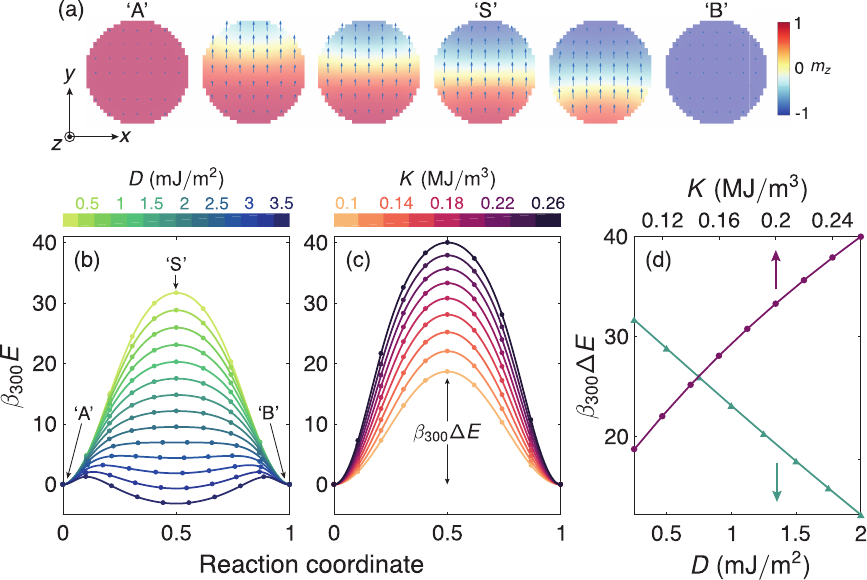}
\caption{(a) Minimum energy path for the magnetization reversal in the disk from the `A' (``up'') to the `B' (``down'') state through the saddle point `S'. (b, c) Energy profile along the reaction coordinate as a function of (b) DMI, $D$, and (c) anisotropy, $K$, with $\beta_{300} \equiv (k_B T_{300})^{-1}$ at $T = 300$ K. (d) Internal energy barriers as a function of $D$ and $K$. \label{fig:mep_and_dE}}
\end{figure}

We begin with the search for minimum energy paths (MEPs) connecting the two stable FM states -- `A' and `B' -- through the energy surface, and the precise identification of the first-order saddle point, `S', along the MEP. This is carried out with the geodesic nudged elastic band method~\cite{gneb}. The MEP is shown in Fig.~\ref{fig:mep_and_dE}a. The reversal takes place via the nucleation of a domain wall at the boundary, which then propagates through the disk. The SP is found as the domain wall reaches the center of the disk, in agreement with previous studies~\cite{khvalkovskiy2013basic,jang2015detrimental, sampaio2016disruptive}. Note that because of the staircase boundary resulting from the discretization, the wall at the SP is not free to rotate around the $z$-axis, and the lowest energy realizations at `S' exist as two perpendicular orientations of the wall, along the $x$- or the $y$-axis. In the absence of dipolar interactions or DMI, there is however a degeneracy between Bloch and N\'eel walls, which manifests in the appearance of a Goldstone eigenmode of zero-energy fluctuations. In reality, dipolar couplings favor Bloch walls in perpendicularly magnetized thin films, and this Goldstone mode has no physical relevance. Therefore, we consistently apply a minimum DMI of 0.25 mJ/m$^3$, which naturally selects N\'eel walls \cite{heide2008dzyaloshinskii, thiaville2012dynamics}, in agreement with Ref.~\cite{sampaio2016disruptive}. We set the effective anisotropy at $K=187$ kJ/m$^3$~\cite{sampaio2016disruptive} and we vary $D$ up to 3.5 mJ/m$^2$.  Fig.~\ref{fig:mep_and_dE}b shows the energy profile along the reaction coordinate for different values of $D$. For $D  \gtrsim 2.5$ mJ/m$^2$, the configuration with the domain wall in the center becomes metastable. In Fig.~\ref{fig:mep_and_dE}d, we show the internal energy barrier as a function of $D$. As one can expect, the energy of the wall varies linearly with $D$ \cite{jang2015detrimental, sampaio2016disruptive, gastaldo2019impact}. We then set $D=0.25$ mJ/m$^2$ and vary $K$ from 0.1 to 0.26 MJ/m$^3$. Below this lower bound, the characteristic wall width, $\delta_w = \sqrt{A/K}$, becomes comparable with the size of the disk and the behavior of the system begins to change. The energy profiles are shown in Fig.~\ref{fig:mep_and_dE}c, and the corresponding energy barriers are shown in Fig. \ref{fig:mep_and_dE}d. Once more, the result matches analytical predictions, as the energy of the wall varies like $\sqrt{K}$.


Next, retention times are calculated as a function of $D$ and $K$ through two distinct approaches. The first relies on the Kramers method~\cite{hanggi1990reaction} following Langer's approach~\cite{langer} for which the rate constant is expressed as
\begin{equation}\label{eq:rate_langer}
k  =\frac{\lambda_+}{2\pi} \Omega_0 e^{-\beta \Delta E} =\frac{\lambda_+}{2\pi} \sqrt{\frac{\prod_i\lambda_i^\mathrm{A}} {\prod_j|\lambda_j^\mathrm{S}|}} e^{-\beta \Delta E}.
\end{equation}
$\lambda_+$ carries the dynamical contribution and corresponds to the rate of growth of the instability at the saddle point. We provide details on its derivation in the Supplemental Material (SM)~\cite{sm_mram}. The $\lambda^{\mathrm{A,S}}$ are the curvatures of the energy surface at `A' and `S'. $\Omega_0$ therefore depends on the details of the fluctuations in the initial state and at the SP and carries the entropic contribution. The theory constitutes the most complete extension of Kramers' method to a multidimensional energy surface, but it is in principle restricted to intermediate to high dampings. In that limit, it can be applied to magnetic spin systems~\cite{coffey,fiedler2012direct} and has been successfully used to calculate the lifetime of magnetic skyrmions~\cite{desplat2018thermal, desplat2019paths, desplat2020path}. At low damping, the density of states deviates more significantly from the equilibrium Maxwell-Boltzmann distribution, so the assumptions of the theory become less valid. Since the typical damping factor for CoFeB thin films is $\alpha~=0.01$~\cite{Bilzer:2006ke}, we compute rates in a range of $\alpha$ from 0.5 to 0.01, while bearing in mind that their accuracy decreases as $\alpha$ decreases. Another important assumption in Kramers' framework is sink boundary conditions past the barrier, i.e., barrier recrossings are not considered. In that sense, the theory should be limited to metastable states decaying to a lower energy minimum. In practice, it can be applied to a symmetric potential~\cite{schratzberger2010validation}, but yields only a forward rate and neglects the recrossings that will necessarily take place. Therefore, the waiting times between reversals are under-evaluated.

Our second approach to rate calculations is forward flux sampling (FFS). This path sampling method, initially developed in the field of biochemistry~\cite{allen2004sampling, allen2006simulating} for simulating rare events and computing activation rates, has also been successfully adapted to magnetic systems to treat problems such as reversal rates in graded media grains~\cite{vogler2013simulating, vogler2015calculating} or collapse rates of skyrmions~\cite{desplat2020path}. In the latter, it yields good agreement with Langer's theory. The method consists in sampling the stochastic magnetization dynamics at finite temperature~\cite{garcia1998langevin} but is significantly more efficient than direct Langevin simulations~\cite{vogler2013simulating} and, as opposed to Langer's theory, does not require any additional assumptions to hold. It is therefore a convenient way to check whether the rate of a particular transition can reasonably fall within the framework of Langer's theory. It relies on a set of interfaces in configuration space, $\{\Lambda_\mathrm{A}, \Lambda_0, \dots \Lambda_n= \Lambda_\mathrm{B}\}$, defined as iso-surfaces of a monotonically varying order parameter, $\zeta$. A flux of trajectories of the system through each interface is computed by Langevin trial runs, which involves integrating the stochastic Landau-Lifshitz equation~\cite{brown1963thermal}. The rate constant of the transition can be decomposed into a product of the partial fluxes as
\begin{equation}
k = \Phi_{\mathrm{A},0} \prod_{i=0}^{n-1} P(\Lambda_{i+1} \vert \Lambda_i),
\label{eq:ffs_rate}
\end{equation}
where $\Phi_{\mathrm{A},0}$ is the rate at which trajectories starting from region `A' cross the first interface $\Lambda_0$, and the conditional probabilities, $P(\Lambda_{i+1} \vert \Lambda_i)$,  correspond to the probability that a trajectory coming from `A' that crossed $\Lambda_i$ for the first time will cross $\Lambda_{i+1}$ before returning to `A'. For best efficiency, FFS requires a pertinent choice of the order parameter. Here we choose the $z$ component of the magnetization averaged over all sites, $\zeta = V^{-1} \int dV \; m_z$.


\begin{figure}
\centering\includegraphics[width=8.5cm]{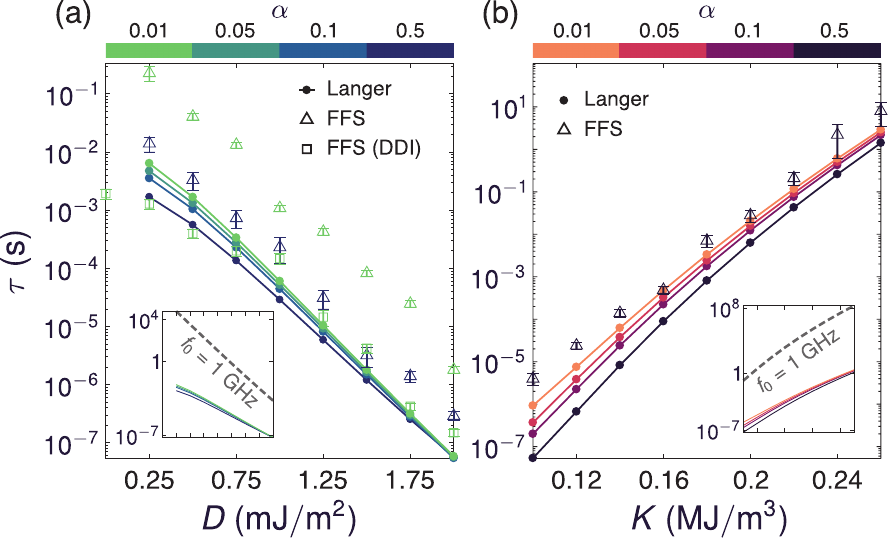}
\caption{Mean waiting time between reversals at 300 K, $\tau$, as a function of the (a) DMI, $D$, and (b) anisotropy, $K$, computed with Langer's theory and FFS for different values of the Gilbert damping, $\alpha$. The insets give a comparison between Langer's theory and the estimate based on $f_0 = 1$ GHz. In (a), a single run with full dipole-dipole interactions (DDI) is also shown.
\label{fig:tau}}
\end{figure}

In Fig.~\ref{fig:tau}, we show the mean waiting time between reversals at $T=300$ K as a function of DMI [Fig. \ref{fig:tau}a ] and anisotropy  [Fig. \ref{fig:tau}b] calculated with a finite-difference implementation of Langer's theory, and FFS simulations performed with a homemade code~\cite{desplat2020path} and MuMax3~\cite{vansteenkiste2014design, Leliaert:2017ci}. With the latter, we performed a single run with full dipole-dipole interactions, which is labeled in Fig.~\ref{fig:tau}(a) as `DDI'. We find a good qualitative agreement between Langer's method and FFS. As expected, FFS yields larger values of $\tau$ than those of Langer's theory, because barrier recrossings are quite frequent. As a result, the flux of trajectories past the SP does not quickly approach unity, like it would if recrossings were negligible. 
When we decrease the damping in FFS, deviations from Langer's theory increase as the assumptions of Langer's become less valid. Micromagnetics simulations are also performed with full dipolar interactions and $\alpha=0.01$. This case is closer to physical systems and is the furthest one from Langer's framework, but the lifetimes are still comparable with Langer's theory. Fig.~\ref{fig:tau}a shows that a $D$ of 1 mJ/m$^2$ leads to a reduction of $\tau$ by about one or two orders of magnitude, and not five-six orders of magnitude like previously anticipated from the constant $f_0$ approximation~\cite{sampaio2016disruptive}. In both cases, the largest stability factors of 32 and 40  [Fig. \ref{fig:mep_and_dE}d] respectively yield a retention time of the order of $10^{-1}-10^{-2}$s, and a few seconds.

\begin{figure}
\centering\includegraphics[width=8.5cm]{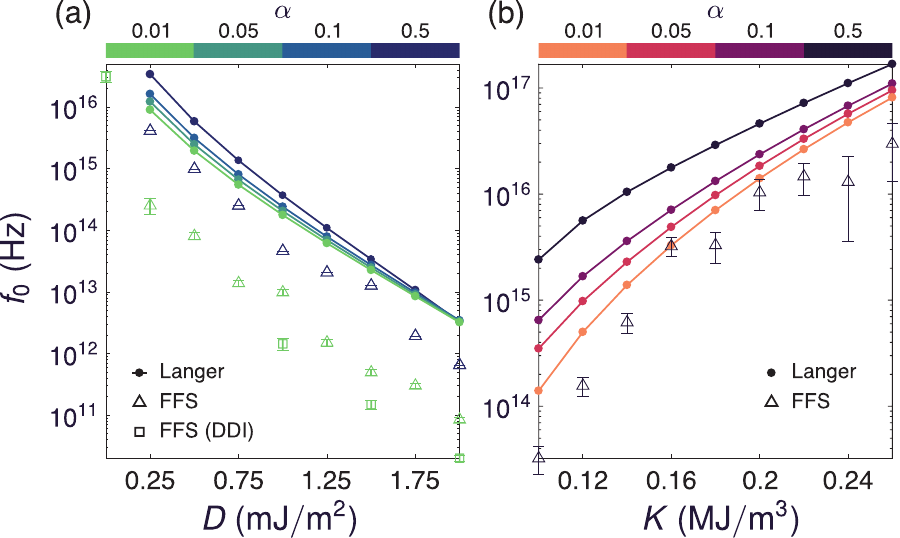}
\caption{Arrhenius prefactor, $f_0$, as a function of the (a) DMI, $D$, and (b) anisotropy, $K$, computed with Langer's theory and FFS for different values of the Gilbert damping, $\alpha$. Points involving full dipole-dipole interactions (DDI) were obtained with energy barriers from Ref.~\onlinecite{sampaio2016disruptive}.
\label{fig:f0}}
\end{figure}

We now examine the variation in the Arrhenius prefactor, which we show in Figs.~\ref{fig:f0}a and \ref{fig:f0}b as a function of $D$ and $K$, respectively. We find values up to $10^{17}$ Hz, which is orders of magnitude greater than the value of 10$^9$ Hz typically  used for estimating thermal stability. These may appear unphysical, but the dynamical contribution, $\lambda_{+}$, does fall in the GHz range, as we show in the SM~\cite{sm_mram}. \emph{The large values of $f_0$ stem from the entropic contribution}. Following previous works \cite{hanggi1990reaction, loxley2006theory, desplat2018thermal, desplat2020path}, we define the change in configurational entropy as
\begin{equation}\label{eq:config_entropy}
e^{\Delta S / k_B} \equiv \sqrt{\frac{\beta}{2\pi}} \sqrt{\frac{\prod_i \lambda_i^A}{\prod_j' \lambda_j^S}},
\end{equation}
where $\prod'$ is defined for positive energy curvatures corresponding to stable modes of fluctuations. It follows that Eq.~(\ref{eq:rate_langer}) can be expressed as
\begin{equation}\label{eq:rate_langer_v2}
k=\frac{\lambda_+}{\sqrt{2\pi \beta |\lambda_1^S|}}e^{\Delta S / k_B} e^{-\beta \Delta E},
\end{equation}
in which $\lambda_1^S$ is the negative curvature at the barrier top and is associated with the unstable mode. The system overcomes the barrier by following this mode -- in our case, it corresponds to the translation of the wall. If $\Delta S$ exhibits a linear dependence on $\Delta E$, we have a case of compensation.

\begin{figure}
\centering\includegraphics[width=6.5cm]{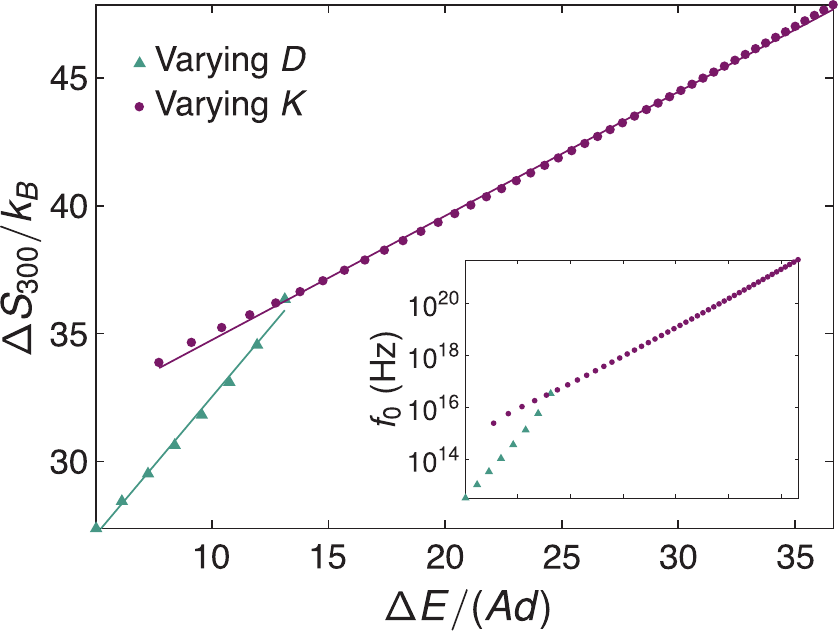}
\caption{Activation entropy at $T$=300 K, $\Delta S_{300}/k_B$, computed with Eq.~(\ref{eq:config_entropy}), as a function of the activation energy normalized by the characteristic exchange energy, $\Delta E/(A d)$, for variations in $D$ and $K$. Solid lines represent linear fits to the data. The inset shows the corresponding variation of $f_0$ for $\alpha=0.5$ with the same horizontal scale.
\label{fig:dS}}
\end{figure}

In Fig. \ref{fig:dS} we plot  $\Delta S/k_B$, as defined in Eq.~(\ref{eq:config_entropy}), as a function of $\Delta E$ for different values of $D$ and $K$. The inset shows the corresponding behavior of the prefactor on a logarithmic scale. We present values of $K$ up to $1$ MJ/m$^3$, for which $f_0$ reaches $10^{21}$ Hz. We reiterate that such values include large entropic contributions, while relaxation processes captured in $\lambda_+$ remain governed by the Landau-Lifshitz equation and lie in the GHz range. In both cases, we find a linear relation between them. The inverse slope is $E_0$~[Eq. (\ref{eq:mn})], the characteristic Meyer-Neldel energy of the family of transitions. Note that $\beta$ does not impact the slope of the graph, and we find compensation regardless of the temperature, as long as Langer's framework remains valid ($\beta \Delta E \ge 5$ \cite{coffey}). When varying $D$, we find $E_0=0.92 A d$, while when varying $K$, we have $E_0=2.06 A d$, in which $Ad$ sets the energy scale of magnons. Ultimately, the activation entropy is found to be more detrimental to the retention time than the DMI.

We offer two ways to interpret this result. First, configurational entropy characterizes the change of volume available to thermal fluctuations on the energy surface induced when the system reaches the SP. Compensation implies that the volume at the saddle point is greater than that of the stable state, i.e., there are more accessible states around the SP, which makes it more likely to be visited. This change of volume increases with the barrier height. This explains the frequent recrossings observed in FFS simulations, and the fact that the system seems to spend a long time in the vicinity of the SP. The magnon dispersion relation in the long wavelength limit behaves like $(M_s/2\gamma) \omega^{\mathrm{A}}(q)= A q^2 + K$ in the FM state, where $q^2 = q_x^2 + q_y^2$ and $\gamma$ is the gyromagnetic constant. For magnons propagating parallel to the wall at the SP, i.e, along $x$, gapless modes appear with the dispersion $(M_s/2\gamma) \omega^{\mathrm{S}}(q_x)= A q_x^2$~\cite{winter1961bloch, garcia2015narrow}. The presence of these gapless, low energy modes at the SP is in line with a largest entropy at the barrier top. In the SM \cite{sm_mram}, we show that the logarithm of the ratio of products of these eigenfrequencies along $x$, i.e., $\ln \prod_i \omega_i^{\mathrm{A}} / \omega_i^{\mathrm{S}}$, can be linked back to $\Omega_0$ in Eq. (\ref{eq:rate_langer}) and tends to behave like $\Delta E$ if $\Delta E \gg Ad$.

A second interpretation lies in the dynamics of the transition. With increasing barrier height, a larger number of small excitations is required to overcome the barrier, and the number of ways to combine these excitations increases~\cite{peacock1982compensation, yelon1990microscopic, yelon1992origin, yelon2006multi}. Compensation therefore results from a multi-excitation process, akin to a ``dynamical entropy'' of the bath~\cite{peacock1982compensation}. The two interpretations are not incompatible, because a large volume around the SP also implies that there must be many pathways on the energy surface that lead to it. Besides the MEP which involves the reaction coordinate, the other paths must necessarily mix the eigencoordinates, which results in the excitation of many magnon modes. It has been suggested that compensation should occur for magnon-driven transitions~\cite{peacock1982compensation}, with a characteristic energy determined by the exchange ($A$) and the energy dependence of the density of states of magnons ($\rho(E) \sim \sqrt{E}$). When varying $A$ at constant $D$ and $K$, we do not find compensation, which seems to confirm this idea. The difference by a factor of two in the slope between $D$- and $K$-driven compensation may stem from the lifting of the degeneracy between counter-propagating azimuthal modes in confined structures with $D$~\cite{GarciaSanchez:2014dw}. 
The authors of Ref.~\onlinecite{peacock1982compensation} suggest that compensation should appear for ensembles of transitions for which $\Delta E$ varies between members, but the effective coupling to the heat bath, remains the same. Compensation may therefore be a general feature of magnetic systems with large activation energies. 


\begin{acknowledgments}
The authors graciously thank Nicolas Reyren for providing published data that we used to compute $f_0$ in Figure~\ref{fig:f0}. This work was supported by the Agence Nationale de la Recherche under Contract No. ANR-17-CE24-0025 (TOPSKY) and the University of Strasbourg Institute for Advanced Study (USIAS) for a Fellowship, within the French national programme ``Investment for the Future'' (IdEx-Unistra).
\end{acknowledgments}

\bibliography{skyrmionbib}

\begin{thebibliography}{48}%
\makeatletter
\providecommand \@ifxundefined [1]{%
 \@ifx{#1\undefined}
}%
\providecommand \@ifnum [1]{%
 \ifnum #1\expandafter \@firstoftwo
 \else \expandafter \@secondoftwo
 \fi
}%
\providecommand \@ifx [1]{%
 \ifx #1\expandafter \@firstoftwo
 \else \expandafter \@secondoftwo
 \fi
}%
\providecommand \natexlab [1]{#1}%
\providecommand \enquote  [1]{``#1''}%
\providecommand \bibnamefont  [1]{#1}%
\providecommand \bibfnamefont [1]{#1}%
\providecommand \citenamefont [1]{#1}%
\providecommand \href@noop [0]{\@secondoftwo}%
\providecommand \href [0]{\begingroup \@sanitize@url \@href}%
\providecommand \@href[1]{\@@startlink{#1}\@@href}%
\providecommand \@@href[1]{\endgroup#1\@@endlink}%
\providecommand \@sanitize@url [0]{\catcode `\\12\catcode `\$12\catcode
  `\&12\catcode `\#12\catcode `\^12\catcode `\_12\catcode `\%12\relax}%
\providecommand \@@startlink[1]{}%
\providecommand \@@endlink[0]{}%
\providecommand \url  [0]{\begingroup\@sanitize@url \@url }%
\providecommand \@url [1]{\endgroup\@href {#1}{\urlprefix }}%
\providecommand \urlprefix  [0]{URL }%
\providecommand \Eprint [0]{\href }%
\providecommand \doibase [0]{http://dx.doi.org/}%
\providecommand \selectlanguage [0]{\@gobble}%
\providecommand \bibinfo  [0]{\@secondoftwo}%
\providecommand \bibfield  [0]{\@secondoftwo}%
\providecommand \translation [1]{[#1]}%
\providecommand \BibitemOpen [0]{}%
\providecommand \bibitemStop [0]{}%
\providecommand \bibitemNoStop [0]{.\EOS\space}%
\providecommand \EOS [0]{\spacefactor3000\relax}%
\providecommand \BibitemShut  [1]{\csname bibitem#1\endcsname}%
\let\auto@bib@innerbib\@empty
\bibitem [{\citenamefont {Constable}(1925)}]{constable1925mechanism}%
  \BibitemOpen
  \bibfield  {author} {\bibinfo {author} {\bibfnamefont {F.~H.}\ \bibnamefont
  {Constable}},\ }\href@noop {} {\bibfield  {journal} {\bibinfo  {journal}
  {Proceedings of the Royal Society of London. Series A, Containing Papers of a
  Mathematical and Physical Character}\ }\textbf {\bibinfo {volume} {108}},\
  \bibinfo {pages} {355} (\bibinfo {year} {1925})}\BibitemShut {NoStop}%
\bibitem [{\citenamefont {Meyer}\ and\ \citenamefont
  {Neldel}(1937)}]{meyer1937relation}%
  \BibitemOpen
  \bibfield  {author} {\bibinfo {author} {\bibfnamefont {W.}~\bibnamefont
  {Meyer}}\ and\ \bibinfo {author} {\bibfnamefont {H.}~\bibnamefont {Neldel}},\
  }\href@noop {} {\bibfield  {journal} {\bibinfo  {journal} {Z. Tech. Phys.}\
  }\textbf {\bibinfo {volume} {18}},\ \bibinfo {pages} {588} (\bibinfo {year}
  {1937})}\BibitemShut {NoStop}%
\bibitem [{\citenamefont {Yelon}\ \emph {et~al.}(2006)\citenamefont {Yelon},
  \citenamefont {Movaghar},\ and\ \citenamefont {Crandall}}]{yelon2006multi}%
  \BibitemOpen
  \bibfield  {author} {\bibinfo {author} {\bibfnamefont {A.}~\bibnamefont
  {Yelon}}, \bibinfo {author} {\bibfnamefont {B.}~\bibnamefont {Movaghar}}, \
  and\ \bibinfo {author} {\bibfnamefont {R.}~\bibnamefont {Crandall}},\
  }\href@noop {} {\bibfield  {journal} {\bibinfo  {journal} {Reports on
  Progress in Physics}\ }\textbf {\bibinfo {volume} {69}},\ \bibinfo {pages}
  {1145} (\bibinfo {year} {2006})}\BibitemShut {NoStop}%
\bibitem [{\citenamefont {Rosenberg}\ \emph {et~al.}(1971)\citenamefont
  {Rosenberg}, \citenamefont {Kemeny}, \citenamefont {Switzer},\ and\
  \citenamefont {Hamilton}}]{rosenberg1971quantitative}%
  \BibitemOpen
  \bibfield  {author} {\bibinfo {author} {\bibfnamefont {B.}~\bibnamefont
  {Rosenberg}}, \bibinfo {author} {\bibfnamefont {G.}~\bibnamefont {Kemeny}},
  \bibinfo {author} {\bibfnamefont {R.~C.}\ \bibnamefont {Switzer}}, \ and\
  \bibinfo {author} {\bibfnamefont {T.~C.}\ \bibnamefont {Hamilton}},\
  }\href@noop {} {\bibfield  {journal} {\bibinfo  {journal} {Nature}\ }\textbf
  {\bibinfo {volume} {232}},\ \bibinfo {pages} {471} (\bibinfo {year}
  {1971})}\BibitemShut {NoStop}%
\bibitem [{\citenamefont {Kamiya}\ \emph {et~al.}(2010)\citenamefont {Kamiya},
  \citenamefont {Nomura},\ and\ \citenamefont {Hosono}}]{kamiya2010present}%
  \BibitemOpen
  \bibfield  {author} {\bibinfo {author} {\bibfnamefont {T.}~\bibnamefont
  {Kamiya}}, \bibinfo {author} {\bibfnamefont {K.}~\bibnamefont {Nomura}}, \
  and\ \bibinfo {author} {\bibfnamefont {H.}~\bibnamefont {Hosono}},\
  }\href@noop {} {\bibfield  {journal} {\bibinfo  {journal} {Science and
  Technology of Advanced Materials}\ }\textbf {\bibinfo {volume} {11}},\
  \bibinfo {pages} {044305} (\bibinfo {year} {2010})}\BibitemShut {NoStop}%
\bibitem [{\citenamefont {Liu}\ and\ \citenamefont
  {Guo}(2001)}]{liu2001isokinetic}%
  \BibitemOpen
  \bibfield  {author} {\bibinfo {author} {\bibfnamefont {L.}~\bibnamefont
  {Liu}}\ and\ \bibinfo {author} {\bibfnamefont {Q.-X.}\ \bibnamefont {Guo}},\
  }\href@noop {} {\bibfield  {journal} {\bibinfo  {journal} {Chemical Reviews}\
  }\textbf {\bibinfo {volume} {101}},\ \bibinfo {pages} {673} (\bibinfo {year}
  {2001})}\BibitemShut {NoStop}%
\bibitem [{\citenamefont {Hart}(1981)}]{hart1981diffusion}%
  \BibitemOpen
  \bibfield  {author} {\bibinfo {author} {\bibfnamefont {S.}~\bibnamefont
  {Hart}},\ }\href@noop {} {\bibfield  {journal} {\bibinfo  {journal}
  {Geochimica et Cosmochimica Acta}\ }\textbf {\bibinfo {volume} {45}},\
  \bibinfo {pages} {279} (\bibinfo {year} {1981})}\BibitemShut {NoStop}%
\bibitem [{\citenamefont {H{\"a}nggi}\ \emph {et~al.}(1990)\citenamefont
  {H{\"a}nggi}, \citenamefont {Talkner},\ and\ \citenamefont
  {Borkovec}}]{hanggi1990reaction}%
  \BibitemOpen
  \bibfield  {author} {\bibinfo {author} {\bibfnamefont {P.}~\bibnamefont
  {H{\"a}nggi}}, \bibinfo {author} {\bibfnamefont {P.}~\bibnamefont {Talkner}},
  \ and\ \bibinfo {author} {\bibfnamefont {M.}~\bibnamefont {Borkovec}},\
  }\href@noop {} {\bibfield  {journal} {\bibinfo  {journal} {Reviews of Modern
  Physics}\ }\textbf {\bibinfo {volume} {62}},\ \bibinfo {pages} {251}
  (\bibinfo {year} {1990})}\BibitemShut {NoStop}%
\bibitem [{\citenamefont {Yelon}\ and\ \citenamefont
  {Movaghar}(1990)}]{yelon1990microscopic}%
  \BibitemOpen
  \bibfield  {author} {\bibinfo {author} {\bibfnamefont {A.}~\bibnamefont
  {Yelon}}\ and\ \bibinfo {author} {\bibfnamefont {B.}~\bibnamefont
  {Movaghar}},\ }\href@noop {} {\bibfield  {journal} {\bibinfo  {journal}
  {Physical Review Letters}\ }\textbf {\bibinfo {volume} {65}},\ \bibinfo
  {pages} {618} (\bibinfo {year} {1990})}\BibitemShut {NoStop}%
\bibitem [{\citenamefont {Yelon}\ \emph {et~al.}(1992)\citenamefont {Yelon},
  \citenamefont {Movaghar},\ and\ \citenamefont {Branz}}]{yelon1992origin}%
  \BibitemOpen
  \bibfield  {author} {\bibinfo {author} {\bibfnamefont {A.}~\bibnamefont
  {Yelon}}, \bibinfo {author} {\bibfnamefont {B.}~\bibnamefont {Movaghar}}, \
  and\ \bibinfo {author} {\bibfnamefont {H.}~\bibnamefont {Branz}},\
  }\href@noop {} {\bibfield  {journal} {\bibinfo  {journal} {Physical Review
  B}\ }\textbf {\bibinfo {volume} {46}},\ \bibinfo {pages} {12244} (\bibinfo
  {year} {1992})}\BibitemShut {NoStop}%
\bibitem [{\citenamefont {Peacock-Lopez}\ and\ \citenamefont
  {Suhl}(1982)}]{peacock1982compensation}%
  \BibitemOpen
  \bibfield  {author} {\bibinfo {author} {\bibfnamefont {E.}~\bibnamefont
  {Peacock-Lopez}}\ and\ \bibinfo {author} {\bibfnamefont {H.}~\bibnamefont
  {Suhl}},\ }\href@noop {} {\bibfield  {journal} {\bibinfo  {journal} {Physical
  Review B}\ }\textbf {\bibinfo {volume} {26}},\ \bibinfo {pages} {3774}
  (\bibinfo {year} {1982})}\BibitemShut {NoStop}%
\bibitem [{\citenamefont {Weller}\ and\ \citenamefont
  {Moser}(1999)}]{weller1999thermal}%
  \BibitemOpen
  \bibfield  {author} {\bibinfo {author} {\bibfnamefont {D.}~\bibnamefont
  {Weller}}\ and\ \bibinfo {author} {\bibfnamefont {A.}~\bibnamefont {Moser}},\
  }\href@noop {} {\bibfield  {journal} {\bibinfo  {journal} {IEEE Transactions
  on Magnetics}\ }\textbf {\bibinfo {volume} {35}},\ \bibinfo {pages} {4423}
  (\bibinfo {year} {1999})}\BibitemShut {NoStop}%
\bibitem [{\citenamefont {Chen}\ \emph {et~al.}(2010)\citenamefont {Chen},
  \citenamefont {Apalkov}, \citenamefont {Diao}, \citenamefont
  {Driskill-Smith}, \citenamefont {Druist}, \citenamefont {Lottis},
  \citenamefont {Nikitin}, \citenamefont {Tang}, \citenamefont {Watts},
  \citenamefont {Wang} \emph {et~al.}}]{chen2010advances}%
  \BibitemOpen
  \bibfield  {author} {\bibinfo {author} {\bibfnamefont {E.}~\bibnamefont
  {Chen}}, \bibinfo {author} {\bibfnamefont {D.}~\bibnamefont {Apalkov}},
  \bibinfo {author} {\bibfnamefont {Z.}~\bibnamefont {Diao}}, \bibinfo {author}
  {\bibfnamefont {A.}~\bibnamefont {Driskill-Smith}}, \bibinfo {author}
  {\bibfnamefont {D.}~\bibnamefont {Druist}}, \bibinfo {author} {\bibfnamefont
  {D.}~\bibnamefont {Lottis}}, \bibinfo {author} {\bibfnamefont
  {V.}~\bibnamefont {Nikitin}}, \bibinfo {author} {\bibfnamefont
  {X.}~\bibnamefont {Tang}}, \bibinfo {author} {\bibfnamefont {S.}~\bibnamefont
  {Watts}}, \bibinfo {author} {\bibfnamefont {S.}~\bibnamefont {Wang}},  \emph
  {et~al.},\ }\href@noop {} {\bibfield  {journal} {\bibinfo  {journal} {IEEE
  Transactions on Magnetics}\ }\textbf {\bibinfo {volume} {46}},\ \bibinfo
  {pages} {1873} (\bibinfo {year} {2010})}\BibitemShut {NoStop}%
\bibitem [{\citenamefont {Lederman}\ \emph {et~al.}(1994)\citenamefont
  {Lederman}, \citenamefont {Schultz},\ and\ \citenamefont
  {Ozaki}}]{lederman1994measurement}%
  \BibitemOpen
  \bibfield  {author} {\bibinfo {author} {\bibfnamefont {M.}~\bibnamefont
  {Lederman}}, \bibinfo {author} {\bibfnamefont {S.}~\bibnamefont {Schultz}}, \
  and\ \bibinfo {author} {\bibfnamefont {M.}~\bibnamefont {Ozaki}},\
  }\href@noop {} {\bibfield  {journal} {\bibinfo  {journal} {Physical Review
  Letters}\ }\textbf {\bibinfo {volume} {73}},\ \bibinfo {pages} {1986}
  (\bibinfo {year} {1994})}\BibitemShut {NoStop}%
\bibitem [{\citenamefont {Cort{\'e}s-Ortu{\~n}o}\ \emph
  {et~al.}(2017)\citenamefont {Cort{\'e}s-Ortu{\~n}o}, \citenamefont {Wang},
  \citenamefont {Beg}, \citenamefont {Pepper}, \citenamefont {Bisotti},
  \citenamefont {Carey}, \citenamefont {Vousden}, \citenamefont {Kluyver},
  \citenamefont {Hovorka},\ and\ \citenamefont {Fangohr}}]{cortes2017thermal}%
  \BibitemOpen
  \bibfield  {author} {\bibinfo {author} {\bibfnamefont {D.}~\bibnamefont
  {Cort{\'e}s-Ortu{\~n}o}}, \bibinfo {author} {\bibfnamefont {W.}~\bibnamefont
  {Wang}}, \bibinfo {author} {\bibfnamefont {M.}~\bibnamefont {Beg}}, \bibinfo
  {author} {\bibfnamefont {R.~A.}\ \bibnamefont {Pepper}}, \bibinfo {author}
  {\bibfnamefont {M.-A.}\ \bibnamefont {Bisotti}}, \bibinfo {author}
  {\bibfnamefont {R.}~\bibnamefont {Carey}}, \bibinfo {author} {\bibfnamefont
  {M.}~\bibnamefont {Vousden}}, \bibinfo {author} {\bibfnamefont
  {T.}~\bibnamefont {Kluyver}}, \bibinfo {author} {\bibfnamefont
  {O.}~\bibnamefont {Hovorka}}, \ and\ \bibinfo {author} {\bibfnamefont
  {H.}~\bibnamefont {Fangohr}},\ }\href@noop {} {\bibfield  {journal} {\bibinfo
   {journal} {Scientific Reports}\ }\textbf {\bibinfo {volume} {7}},\ \bibinfo
  {pages} {4060} (\bibinfo {year} {2017})}\BibitemShut {NoStop}%
\bibitem [{\citenamefont {Khvalkovskiy}\ \emph {et~al.}(2013)\citenamefont
  {Khvalkovskiy}, \citenamefont {Apalkov}, \citenamefont {Watts}, \citenamefont
  {Chepulskii}, \citenamefont {Beach}, \citenamefont {Ong}, \citenamefont
  {Tang}, \citenamefont {Driskill-Smith}, \citenamefont {Butler}, \citenamefont
  {Visscher} \emph {et~al.}}]{khvalkovskiy2013basic}%
  \BibitemOpen
  \bibfield  {author} {\bibinfo {author} {\bibfnamefont {A.}~\bibnamefont
  {Khvalkovskiy}}, \bibinfo {author} {\bibfnamefont {D.}~\bibnamefont
  {Apalkov}}, \bibinfo {author} {\bibfnamefont {S.}~\bibnamefont {Watts}},
  \bibinfo {author} {\bibfnamefont {R.}~\bibnamefont {Chepulskii}}, \bibinfo
  {author} {\bibfnamefont {R.}~\bibnamefont {Beach}}, \bibinfo {author}
  {\bibfnamefont {A.}~\bibnamefont {Ong}}, \bibinfo {author} {\bibfnamefont
  {X.}~\bibnamefont {Tang}}, \bibinfo {author} {\bibfnamefont {A.}~\bibnamefont
  {Driskill-Smith}}, \bibinfo {author} {\bibfnamefont {W.}~\bibnamefont
  {Butler}}, \bibinfo {author} {\bibfnamefont {P.}~\bibnamefont {Visscher}},
  \emph {et~al.},\ }\href@noop {} {\bibfield  {journal} {\bibinfo  {journal}
  {Journal of Physics D: Applied Physics}\ }\textbf {\bibinfo {volume} {46}},\
  \bibinfo {pages} {074001} (\bibinfo {year} {2013})}\BibitemShut {NoStop}%
\bibitem [{\citenamefont {Wild}\ \emph {et~al.}(2017)\citenamefont {Wild},
  \citenamefont {Meier}, \citenamefont {P{\"o}llath}, \citenamefont
  {Kronseder}, \citenamefont {Bauer}, \citenamefont {Chacon}, \citenamefont
  {Halder}, \citenamefont {Schowalter}, \citenamefont {Rosenauer},
  \citenamefont {Zweck} \emph {et~al.}}]{wild2017entropy}%
  \BibitemOpen
  \bibfield  {author} {\bibinfo {author} {\bibfnamefont {J.}~\bibnamefont
  {Wild}}, \bibinfo {author} {\bibfnamefont {T.~N.}\ \bibnamefont {Meier}},
  \bibinfo {author} {\bibfnamefont {S.}~\bibnamefont {P{\"o}llath}}, \bibinfo
  {author} {\bibfnamefont {M.}~\bibnamefont {Kronseder}}, \bibinfo {author}
  {\bibfnamefont {A.}~\bibnamefont {Bauer}}, \bibinfo {author} {\bibfnamefont
  {A.}~\bibnamefont {Chacon}}, \bibinfo {author} {\bibfnamefont
  {M.}~\bibnamefont {Halder}}, \bibinfo {author} {\bibfnamefont
  {M.}~\bibnamefont {Schowalter}}, \bibinfo {author} {\bibfnamefont
  {A.}~\bibnamefont {Rosenauer}}, \bibinfo {author} {\bibfnamefont
  {J.}~\bibnamefont {Zweck}},  \emph {et~al.},\ }\href@noop {} {\bibfield
  {journal} {\bibinfo  {journal} {Science Advances}\ }\textbf {\bibinfo
  {volume} {3}},\ \bibinfo {pages} {e1701704} (\bibinfo {year}
  {2017})}\BibitemShut {NoStop}%
\bibitem [{\citenamefont {Bessarab}\ \emph {et~al.}(2018)\citenamefont
  {Bessarab}, \citenamefont {M{\"u}ller}, \citenamefont {Lobanov},
  \citenamefont {Rybakov}, \citenamefont {Kiselev}, \citenamefont
  {J{\'o}nsson}, \citenamefont {Uzdin}, \citenamefont {Bl{\"u}gel},
  \citenamefont {Bergqvist},\ and\ \citenamefont
  {Delin}}]{bessarab2018annihilation}%
  \BibitemOpen
  \bibfield  {author} {\bibinfo {author} {\bibfnamefont {P.~F.}\ \bibnamefont
  {Bessarab}}, \bibinfo {author} {\bibfnamefont {G.~P.}\ \bibnamefont
  {M{\"u}ller}}, \bibinfo {author} {\bibfnamefont {I.~S.}\ \bibnamefont
  {Lobanov}}, \bibinfo {author} {\bibfnamefont {F.~N.}\ \bibnamefont
  {Rybakov}}, \bibinfo {author} {\bibfnamefont {N.~S.}\ \bibnamefont
  {Kiselev}}, \bibinfo {author} {\bibfnamefont {H.}~\bibnamefont
  {J{\'o}nsson}}, \bibinfo {author} {\bibfnamefont {V.~M.}\ \bibnamefont
  {Uzdin}}, \bibinfo {author} {\bibfnamefont {S.}~\bibnamefont {Bl{\"u}gel}},
  \bibinfo {author} {\bibfnamefont {L.}~\bibnamefont {Bergqvist}}, \ and\
  \bibinfo {author} {\bibfnamefont {A.}~\bibnamefont {Delin}},\ }\href@noop {}
  {\bibfield  {journal} {\bibinfo  {journal} {Scientific Reports}\ }\textbf
  {\bibinfo {volume} {8}},\ \bibinfo {pages} {618} (\bibinfo {year}
  {2018})}\BibitemShut {NoStop}%
\bibitem [{\citenamefont {Desplat}\ \emph {et~al.}(2018)\citenamefont
  {Desplat}, \citenamefont {Suess}, \citenamefont {Kim},\ and\ \citenamefont
  {Stamps}}]{desplat2018thermal}%
  \BibitemOpen
  \bibfield  {author} {\bibinfo {author} {\bibfnamefont {L.}~\bibnamefont
  {Desplat}}, \bibinfo {author} {\bibfnamefont {D.}~\bibnamefont {Suess}},
  \bibinfo {author} {\bibfnamefont {J.-V.}\ \bibnamefont {Kim}}, \ and\
  \bibinfo {author} {\bibfnamefont {R.~L.}\ \bibnamefont {Stamps}},\
  }\href@noop {} {\bibfield  {journal} {\bibinfo  {journal} {Physical Review
  B}\ }\textbf {\bibinfo {volume} {98}},\ \bibinfo {pages} {134407} (\bibinfo
  {year} {2018})}\BibitemShut {NoStop}%
\bibitem [{\citenamefont {von Malottki}\ \emph {et~al.}(2019)\citenamefont {von
  Malottki}, \citenamefont {Bessarab}, \citenamefont {Haldar}, \citenamefont
  {Delin},\ and\ \citenamefont {Heinze}}]{von2019skyrmion}%
  \BibitemOpen
  \bibfield  {author} {\bibinfo {author} {\bibfnamefont {S.}~\bibnamefont {von
  Malottki}}, \bibinfo {author} {\bibfnamefont {P.~F.}\ \bibnamefont
  {Bessarab}}, \bibinfo {author} {\bibfnamefont {S.}~\bibnamefont {Haldar}},
  \bibinfo {author} {\bibfnamefont {A.}~\bibnamefont {Delin}}, \ and\ \bibinfo
  {author} {\bibfnamefont {S.}~\bibnamefont {Heinze}},\ }\href@noop {}
  {\bibfield  {journal} {\bibinfo  {journal} {Physical Review B}\ }\textbf
  {\bibinfo {volume} {99}},\ \bibinfo {pages} {060409} (\bibinfo {year}
  {2019})}\BibitemShut {NoStop}%
\bibitem [{\citenamefont {Desplat}\ \emph {et~al.}(2020)\citenamefont
  {Desplat}, \citenamefont {Vogler}, \citenamefont {Kim}, \citenamefont
  {Stamps},\ and\ \citenamefont {Suess}}]{desplat2020path}%
  \BibitemOpen
  \bibfield  {author} {\bibinfo {author} {\bibfnamefont {L.}~\bibnamefont
  {Desplat}}, \bibinfo {author} {\bibfnamefont {C.}~\bibnamefont {Vogler}},
  \bibinfo {author} {\bibfnamefont {J.-V.}\ \bibnamefont {Kim}}, \bibinfo
  {author} {\bibfnamefont {R.~L.}\ \bibnamefont {Stamps}}, \ and\ \bibinfo
  {author} {\bibfnamefont {D.}~\bibnamefont {Suess}},\ }\href@noop {}
  {\bibfield  {journal} {\bibinfo  {journal} {Physical Review B}\ }\textbf
  {\bibinfo {volume} {101}},\ \bibinfo {pages} {060403} (\bibinfo {year}
  {2020})}\BibitemShut {NoStop}%
\bibitem [{\citenamefont {Heinze}\ \emph {et~al.}(2011)\citenamefont {Heinze},
  \citenamefont {Von~Bergmann}, \citenamefont {Menzel}, \citenamefont {Brede},
  \citenamefont {Kubetzka}, \citenamefont {Wiesendanger}, \citenamefont
  {Bihlmayer},\ and\ \citenamefont {Bl{\"u}gel}}]{heinze2011spontaneous}%
  \BibitemOpen
  \bibfield  {author} {\bibinfo {author} {\bibfnamefont {S.}~\bibnamefont
  {Heinze}}, \bibinfo {author} {\bibfnamefont {K.}~\bibnamefont
  {Von~Bergmann}}, \bibinfo {author} {\bibfnamefont {M.}~\bibnamefont
  {Menzel}}, \bibinfo {author} {\bibfnamefont {J.}~\bibnamefont {Brede}},
  \bibinfo {author} {\bibfnamefont {A.}~\bibnamefont {Kubetzka}}, \bibinfo
  {author} {\bibfnamefont {R.}~\bibnamefont {Wiesendanger}}, \bibinfo {author}
  {\bibfnamefont {G.}~\bibnamefont {Bihlmayer}}, \ and\ \bibinfo {author}
  {\bibfnamefont {S.}~\bibnamefont {Bl{\"u}gel}},\ }\href@noop {} {\bibfield
  {journal} {\bibinfo  {journal} {Nature Physics}\ }\textbf {\bibinfo {volume}
  {7}},\ \bibinfo {pages} {713} (\bibinfo {year} {2011})}\BibitemShut {NoStop}%
\bibitem [{\citenamefont {Sampaio}\ \emph {et~al.}(2016)\citenamefont
  {Sampaio}, \citenamefont {Khvalkovskiy}, \citenamefont {Kuteifan},
  \citenamefont {Cubukcu}, \citenamefont {Apalkov}, \citenamefont {Lomakin},
  \citenamefont {Cros},\ and\ \citenamefont {Reyren}}]{sampaio2016disruptive}%
  \BibitemOpen
  \bibfield  {author} {\bibinfo {author} {\bibfnamefont {J.}~\bibnamefont
  {Sampaio}}, \bibinfo {author} {\bibfnamefont {A.}~\bibnamefont
  {Khvalkovskiy}}, \bibinfo {author} {\bibfnamefont {M.}~\bibnamefont
  {Kuteifan}}, \bibinfo {author} {\bibfnamefont {M.}~\bibnamefont {Cubukcu}},
  \bibinfo {author} {\bibfnamefont {D.}~\bibnamefont {Apalkov}}, \bibinfo
  {author} {\bibfnamefont {V.}~\bibnamefont {Lomakin}}, \bibinfo {author}
  {\bibfnamefont {V.}~\bibnamefont {Cros}}, \ and\ \bibinfo {author}
  {\bibfnamefont {N.}~\bibnamefont {Reyren}},\ }\href@noop {} {\bibfield
  {journal} {\bibinfo  {journal} {Applied Physics Letters}\ }\textbf {\bibinfo
  {volume} {108}},\ \bibinfo {pages} {112403} (\bibinfo {year}
  {2016})}\BibitemShut {NoStop}%
\bibitem [{\citenamefont {Gastaldo}\ \emph {et~al.}(2019)\citenamefont
  {Gastaldo}, \citenamefont {Strelkov}, \citenamefont {Buda-Prejbeanu},
  \citenamefont {Dieny}, \citenamefont {Boulle}, \citenamefont {Allia},\ and\
  \citenamefont {Tiberto}}]{gastaldo2019impact}%
  \BibitemOpen
  \bibfield  {author} {\bibinfo {author} {\bibfnamefont {D.}~\bibnamefont
  {Gastaldo}}, \bibinfo {author} {\bibfnamefont {N.}~\bibnamefont {Strelkov}},
  \bibinfo {author} {\bibfnamefont {L.~D.}\ \bibnamefont {Buda-Prejbeanu}},
  \bibinfo {author} {\bibfnamefont {B.}~\bibnamefont {Dieny}}, \bibinfo
  {author} {\bibfnamefont {O.}~\bibnamefont {Boulle}}, \bibinfo {author}
  {\bibfnamefont {P.}~\bibnamefont {Allia}}, \ and\ \bibinfo {author}
  {\bibfnamefont {P.}~\bibnamefont {Tiberto}},\ }\href@noop {} {\bibfield
  {journal} {\bibinfo  {journal} {Journal of Applied Physics}\ }\textbf
  {\bibinfo {volume} {126}},\ \bibinfo {pages} {103905} (\bibinfo {year}
  {2019})}\BibitemShut {NoStop}%
\bibitem [{\citenamefont {Langer}(1969)}]{langer}%
  \BibitemOpen
  \bibfield  {author} {\bibinfo {author} {\bibfnamefont {J.~S.}\ \bibnamefont
  {Langer}},\ }\href@noop {} {\bibfield  {journal} {\bibinfo  {journal} {Annals
  of Physics}\ }\textbf {\bibinfo {volume} {54}},\ \bibinfo {pages} {258}
  (\bibinfo {year} {1969})}\BibitemShut {NoStop}%
\bibitem [{\citenamefont {Allen}\ \emph {et~al.}(2005)\citenamefont {Allen},
  \citenamefont {Warren},\ and\ \citenamefont {ten Wolde}}]{allen2004sampling}%
  \BibitemOpen
  \bibfield  {author} {\bibinfo {author} {\bibfnamefont {R.~J.}\ \bibnamefont
  {Allen}}, \bibinfo {author} {\bibfnamefont {P.~B.}\ \bibnamefont {Warren}}, \
  and\ \bibinfo {author} {\bibfnamefont {P.~R.}\ \bibnamefont {ten Wolde}},\
  }\href@noop {} {\bibfield  {journal} {\bibinfo  {journal} {Physical Review
  Letters}\ }\textbf {\bibinfo {volume} {94}},\ \bibinfo {pages} {018104}
  (\bibinfo {year} {2005})}\BibitemShut {NoStop}%
\bibitem [{\citenamefont {Allen}\ \emph {et~al.}(2006)\citenamefont {Allen},
  \citenamefont {Frenkel},\ and\ \citenamefont {ten
  Wolde}}]{allen2006simulating}%
  \BibitemOpen
  \bibfield  {author} {\bibinfo {author} {\bibfnamefont {R.~J.}\ \bibnamefont
  {Allen}}, \bibinfo {author} {\bibfnamefont {D.}~\bibnamefont {Frenkel}}, \
  and\ \bibinfo {author} {\bibfnamefont {P.~R.}\ \bibnamefont {ten Wolde}},\
  }\href@noop {} {\bibfield  {journal} {\bibinfo  {journal} {The Journal of
  Chemical Physics}\ }\textbf {\bibinfo {volume} {124}},\ \bibinfo {pages}
  {024102} (\bibinfo {year} {2006})}\BibitemShut {NoStop}%
\bibitem [{\citenamefont {Chen}\ \emph {et~al.}(1991)\citenamefont {Chen},
  \citenamefont {Brug},\ and\ \citenamefont
  {Goldfarb}}]{chen1991demagnetizing}%
  \BibitemOpen
  \bibfield  {author} {\bibinfo {author} {\bibfnamefont {D.-X.}\ \bibnamefont
  {Chen}}, \bibinfo {author} {\bibfnamefont {J.~A.}\ \bibnamefont {Brug}}, \
  and\ \bibinfo {author} {\bibfnamefont {R.~B.}\ \bibnamefont {Goldfarb}},\
  }\href@noop {} {\bibfield  {journal} {\bibinfo  {journal} {IEEE Transactions
  on magnetics}\ }\textbf {\bibinfo {volume} {27}},\ \bibinfo {pages} {3601}
  (\bibinfo {year} {1991})}\BibitemShut {NoStop}%
\bibitem [{\citenamefont {Bessarab}\ \emph {et~al.}(2015)\citenamefont
  {Bessarab}, \citenamefont {Uzdin},\ and\ \citenamefont {Jonsson}}]{gneb}%
  \BibitemOpen
  \bibfield  {author} {\bibinfo {author} {\bibfnamefont {P.~F.}\ \bibnamefont
  {Bessarab}}, \bibinfo {author} {\bibfnamefont {V.~M.}\ \bibnamefont {Uzdin}},
  \ and\ \bibinfo {author} {\bibfnamefont {H.}~\bibnamefont {Jonsson}},\
  }\href@noop {} {\bibfield  {journal} {\bibinfo  {journal} {Computer Physics
  Communications}\ }\textbf {\bibinfo {volume} {196}},\ \bibinfo {pages} {335}
  (\bibinfo {year} {2015})}\BibitemShut {NoStop}%
\bibitem [{\citenamefont {Jang}\ \emph {et~al.}(2015)\citenamefont {Jang},
  \citenamefont {Song}, \citenamefont {Lee}, \citenamefont {Lee},\ and\
  \citenamefont {Lee}}]{jang2015detrimental}%
  \BibitemOpen
  \bibfield  {author} {\bibinfo {author} {\bibfnamefont {P.-H.}\ \bibnamefont
  {Jang}}, \bibinfo {author} {\bibfnamefont {K.}~\bibnamefont {Song}}, \bibinfo
  {author} {\bibfnamefont {S.-J.}\ \bibnamefont {Lee}}, \bibinfo {author}
  {\bibfnamefont {S.-W.}\ \bibnamefont {Lee}}, \ and\ \bibinfo {author}
  {\bibfnamefont {K.-J.}\ \bibnamefont {Lee}},\ }\href@noop {} {\bibfield
  {journal} {\bibinfo  {journal} {Applied Physics Letters}\ }\textbf {\bibinfo
  {volume} {107}},\ \bibinfo {pages} {202401} (\bibinfo {year}
  {2015})}\BibitemShut {NoStop}%
\bibitem [{\citenamefont {Heide}\ \emph {et~al.}(2008)\citenamefont {Heide},
  \citenamefont {Bihlmayer},\ and\ \citenamefont
  {Bl{\"u}gel}}]{heide2008dzyaloshinskii}%
  \BibitemOpen
  \bibfield  {author} {\bibinfo {author} {\bibfnamefont {M.}~\bibnamefont
  {Heide}}, \bibinfo {author} {\bibfnamefont {G.}~\bibnamefont {Bihlmayer}}, \
  and\ \bibinfo {author} {\bibfnamefont {S.}~\bibnamefont {Bl{\"u}gel}},\
  }\href@noop {} {\bibfield  {journal} {\bibinfo  {journal} {Physical Review
  B}\ }\textbf {\bibinfo {volume} {78}},\ \bibinfo {pages} {140403} (\bibinfo
  {year} {2008})}\BibitemShut {NoStop}%
\bibitem [{\citenamefont {Thiaville}\ \emph {et~al.}(2012)\citenamefont
  {Thiaville}, \citenamefont {Rohart}, \citenamefont {Ju{\'e}}, \citenamefont
  {Cros},\ and\ \citenamefont {Fert}}]{thiaville2012dynamics}%
  \BibitemOpen
  \bibfield  {author} {\bibinfo {author} {\bibfnamefont {A.}~\bibnamefont
  {Thiaville}}, \bibinfo {author} {\bibfnamefont {S.}~\bibnamefont {Rohart}},
  \bibinfo {author} {\bibfnamefont {{\'E}.}~\bibnamefont {Ju{\'e}}}, \bibinfo
  {author} {\bibfnamefont {V.}~\bibnamefont {Cros}}, \ and\ \bibinfo {author}
  {\bibfnamefont {A.}~\bibnamefont {Fert}},\ }\href@noop {} {\bibfield
  {journal} {\bibinfo  {journal} {Europhysics Letters}\ }\textbf {\bibinfo
  {volume} {100}},\ \bibinfo {pages} {57002} (\bibinfo {year}
  {2012})}\BibitemShut {NoStop}%
\bibitem [{sm_()}]{sm_mram}%
  \BibitemOpen
  \href@noop {} {}\bibinfo {note} {See Supplemental Material at [URL] for
  additional information.}\BibitemShut {Stop}%
\bibitem [{\citenamefont {Coffey}\ \emph {et~al.}(2001)\citenamefont {Coffey},
  \citenamefont {Garanin},\ and\ \citenamefont {McCarthy}}]{coffey}%
  \BibitemOpen
  \bibfield  {author} {\bibinfo {author} {\bibfnamefont {W.}~\bibnamefont
  {Coffey}}, \bibinfo {author} {\bibfnamefont {D.}~\bibnamefont {Garanin}}, \
  and\ \bibinfo {author} {\bibfnamefont {D.}~\bibnamefont {McCarthy}},\
  }\href@noop {} {\bibfield  {journal} {\bibinfo  {journal} {Advances in
  Chemical Physics}\ }\textbf {\bibinfo {volume} {117}},\ \bibinfo {pages}
  {483} (\bibinfo {year} {2001})}\BibitemShut {NoStop}%
\bibitem [{\citenamefont {Fiedler}\ \emph {et~al.}(2012)\citenamefont
  {Fiedler}, \citenamefont {Fidler}, \citenamefont {Lee}, \citenamefont
  {Schrefl}, \citenamefont {Stamps}, \citenamefont {Braun},\ and\ \citenamefont
  {Suess}}]{fiedler2012direct}%
  \BibitemOpen
  \bibfield  {author} {\bibinfo {author} {\bibfnamefont {G.}~\bibnamefont
  {Fiedler}}, \bibinfo {author} {\bibfnamefont {J.}~\bibnamefont {Fidler}},
  \bibinfo {author} {\bibfnamefont {J.}~\bibnamefont {Lee}}, \bibinfo {author}
  {\bibfnamefont {T.}~\bibnamefont {Schrefl}}, \bibinfo {author} {\bibfnamefont
  {R.~L.}\ \bibnamefont {Stamps}}, \bibinfo {author} {\bibfnamefont
  {H.}~\bibnamefont {Braun}}, \ and\ \bibinfo {author} {\bibfnamefont
  {D.}~\bibnamefont {Suess}},\ }\href@noop {} {\bibfield  {journal} {\bibinfo
  {journal} {Journal of Applied Physics}\ }\textbf {\bibinfo {volume} {111}},\
  \bibinfo {pages} {093917} (\bibinfo {year} {2012})}\BibitemShut {NoStop}%
\bibitem [{\citenamefont {Desplat}\ \emph {et~al.}(2019)\citenamefont
  {Desplat}, \citenamefont {Kim},\ and\ \citenamefont
  {Stamps}}]{desplat2019paths}%
  \BibitemOpen
  \bibfield  {author} {\bibinfo {author} {\bibfnamefont {L.}~\bibnamefont
  {Desplat}}, \bibinfo {author} {\bibfnamefont {J.-V.}\ \bibnamefont {Kim}}, \
  and\ \bibinfo {author} {\bibfnamefont {R.~L.}\ \bibnamefont {Stamps}},\
  }\href@noop {} {\bibfield  {journal} {\bibinfo  {journal} {Physical Review
  B}\ }\textbf {\bibinfo {volume} {99}},\ \bibinfo {pages} {174409} (\bibinfo
  {year} {2019})}\BibitemShut {NoStop}%
\bibitem [{\citenamefont {Bilzer}\ \emph {et~al.}(2006)\citenamefont {Bilzer},
  \citenamefont {Devolder}, \citenamefont {Kim}, \citenamefont {Counil},
  \citenamefont {Chappert}, \citenamefont {Cardoso},\ and\ \citenamefont
  {Freitas}}]{Bilzer:2006ke}%
  \BibitemOpen
  \bibfield  {author} {\bibinfo {author} {\bibfnamefont {C.}~\bibnamefont
  {Bilzer}}, \bibinfo {author} {\bibfnamefont {T.}~\bibnamefont {Devolder}},
  \bibinfo {author} {\bibfnamefont {J.-V.}\ \bibnamefont {Kim}}, \bibinfo
  {author} {\bibfnamefont {G.}~\bibnamefont {Counil}}, \bibinfo {author}
  {\bibfnamefont {C.}~\bibnamefont {Chappert}}, \bibinfo {author}
  {\bibfnamefont {S.}~\bibnamefont {Cardoso}}, \ and\ \bibinfo {author}
  {\bibfnamefont {P.~P.}\ \bibnamefont {Freitas}},\ }\href@noop {} {\bibfield
  {journal} {\bibinfo  {journal} {Journal of Applied Physics}\ }\textbf
  {\bibinfo {volume} {100}},\ \bibinfo {pages} {053903} (\bibinfo {year}
  {2006})}\BibitemShut {NoStop}%
\bibitem [{\citenamefont {Schratzberger}\ \emph {et~al.}(2010)\citenamefont
  {Schratzberger}, \citenamefont {Lee}, \citenamefont {Fuger}, \citenamefont
  {Fidler}, \citenamefont {Fiedler}, \citenamefont {Schrefl},\ and\
  \citenamefont {Suess}}]{schratzberger2010validation}%
  \BibitemOpen
  \bibfield  {author} {\bibinfo {author} {\bibfnamefont {J.}~\bibnamefont
  {Schratzberger}}, \bibinfo {author} {\bibfnamefont {J.}~\bibnamefont {Lee}},
  \bibinfo {author} {\bibfnamefont {M.}~\bibnamefont {Fuger}}, \bibinfo
  {author} {\bibfnamefont {J.}~\bibnamefont {Fidler}}, \bibinfo {author}
  {\bibfnamefont {G.}~\bibnamefont {Fiedler}}, \bibinfo {author} {\bibfnamefont
  {T.}~\bibnamefont {Schrefl}}, \ and\ \bibinfo {author} {\bibfnamefont
  {D.}~\bibnamefont {Suess}},\ }\href@noop {} {\bibfield  {journal} {\bibinfo
  {journal} {Journal of Applied Physics}\ }\textbf {\bibinfo {volume} {108}},\
  \bibinfo {pages} {033915} (\bibinfo {year} {2010})}\BibitemShut {NoStop}%
\bibitem [{\citenamefont {Vogler}\ \emph {et~al.}(2013)\citenamefont {Vogler},
  \citenamefont {Bruckner}, \citenamefont {Bergmair}, \citenamefont {Huber},
  \citenamefont {Suess},\ and\ \citenamefont {Dellago}}]{vogler2013simulating}%
  \BibitemOpen
  \bibfield  {author} {\bibinfo {author} {\bibfnamefont {C.}~\bibnamefont
  {Vogler}}, \bibinfo {author} {\bibfnamefont {F.}~\bibnamefont {Bruckner}},
  \bibinfo {author} {\bibfnamefont {B.}~\bibnamefont {Bergmair}}, \bibinfo
  {author} {\bibfnamefont {T.}~\bibnamefont {Huber}}, \bibinfo {author}
  {\bibfnamefont {D.}~\bibnamefont {Suess}}, \ and\ \bibinfo {author}
  {\bibfnamefont {C.}~\bibnamefont {Dellago}},\ }\href@noop {} {\bibfield
  {journal} {\bibinfo  {journal} {Physical Review B}\ }\textbf {\bibinfo
  {volume} {88}},\ \bibinfo {pages} {134409} (\bibinfo {year}
  {2013})}\BibitemShut {NoStop}%
\bibitem [{\citenamefont {Vogler}\ \emph {et~al.}(2015)\citenamefont {Vogler},
  \citenamefont {Bruckner}, \citenamefont {Suess},\ and\ \citenamefont
  {Dellago}}]{vogler2015calculating}%
  \BibitemOpen
  \bibfield  {author} {\bibinfo {author} {\bibfnamefont {C.}~\bibnamefont
  {Vogler}}, \bibinfo {author} {\bibfnamefont {F.}~\bibnamefont {Bruckner}},
  \bibinfo {author} {\bibfnamefont {D.}~\bibnamefont {Suess}}, \ and\ \bibinfo
  {author} {\bibfnamefont {C.}~\bibnamefont {Dellago}},\ }\href@noop {}
  {\bibfield  {journal} {\bibinfo  {journal} {Journal of Applied Physics}\
  }\textbf {\bibinfo {volume} {117}},\ \bibinfo {pages} {163907} (\bibinfo
  {year} {2015})}\BibitemShut {NoStop}%
\bibitem [{\citenamefont {Garc{\'\i}a-Palacios}\ and\ \citenamefont
  {L{\'a}zaro}(1998)}]{garcia1998langevin}%
  \BibitemOpen
  \bibfield  {author} {\bibinfo {author} {\bibfnamefont {J.~L.}\ \bibnamefont
  {Garc{\'\i}a-Palacios}}\ and\ \bibinfo {author} {\bibfnamefont {F.~J.}\
  \bibnamefont {L{\'a}zaro}},\ }\href@noop {} {\bibfield  {journal} {\bibinfo
  {journal} {Physical Review B}\ }\textbf {\bibinfo {volume} {58}},\ \bibinfo
  {pages} {14937} (\bibinfo {year} {1998})}\BibitemShut {NoStop}%
\bibitem [{\citenamefont {Brown~Jr}(1963)}]{brown1963thermal}%
  \BibitemOpen
  \bibfield  {author} {\bibinfo {author} {\bibfnamefont {W.~F.}\ \bibnamefont
  {Brown~Jr}},\ }\href@noop {} {\bibfield  {journal} {\bibinfo  {journal}
  {Physical Review}\ }\textbf {\bibinfo {volume} {130}},\ \bibinfo {pages}
  {1677} (\bibinfo {year} {1963})}\BibitemShut {NoStop}%
\bibitem [{\citenamefont {Vansteenkiste}\ \emph {et~al.}(2014)\citenamefont
  {Vansteenkiste}, \citenamefont {Leliaert}, \citenamefont {Dvornik},
  \citenamefont {Helsen}, \citenamefont {Garcia-Sanchez},\ and\ \citenamefont
  {Van~Waeyenberge}}]{vansteenkiste2014design}%
  \BibitemOpen
  \bibfield  {author} {\bibinfo {author} {\bibfnamefont {A.}~\bibnamefont
  {Vansteenkiste}}, \bibinfo {author} {\bibfnamefont {J.}~\bibnamefont
  {Leliaert}}, \bibinfo {author} {\bibfnamefont {M.}~\bibnamefont {Dvornik}},
  \bibinfo {author} {\bibfnamefont {M.}~\bibnamefont {Helsen}}, \bibinfo
  {author} {\bibfnamefont {F.}~\bibnamefont {Garcia-Sanchez}}, \ and\ \bibinfo
  {author} {\bibfnamefont {B.}~\bibnamefont {Van~Waeyenberge}},\ }\href@noop {}
  {\bibfield  {journal} {\bibinfo  {journal} {AIP Advances}\ }\textbf {\bibinfo
  {volume} {4}},\ \bibinfo {pages} {107133} (\bibinfo {year}
  {2014})}\BibitemShut {NoStop}%
\bibitem [{\citenamefont {Leliaert}\ \emph {et~al.}(2017)\citenamefont
  {Leliaert}, \citenamefont {Mulkers}, \citenamefont {De~Clercq}, \citenamefont
  {Coene}, \citenamefont {Dvornik},\ and\ \citenamefont
  {Van~Waeyenberge}}]{Leliaert:2017ci}%
  \BibitemOpen
  \bibfield  {author} {\bibinfo {author} {\bibfnamefont {J.}~\bibnamefont
  {Leliaert}}, \bibinfo {author} {\bibfnamefont {J.}~\bibnamefont {Mulkers}},
  \bibinfo {author} {\bibfnamefont {J.}~\bibnamefont {De~Clercq}}, \bibinfo
  {author} {\bibfnamefont {A.}~\bibnamefont {Coene}}, \bibinfo {author}
  {\bibfnamefont {M.}~\bibnamefont {Dvornik}}, \ and\ \bibinfo {author}
  {\bibfnamefont {B.}~\bibnamefont {Van~Waeyenberge}},\ }\href@noop {}
  {\bibfield  {journal} {\bibinfo  {journal} {AIP Advances}\ }\textbf {\bibinfo
  {volume} {7}},\ \bibinfo {pages} {125010} (\bibinfo {year}
  {2017})}\BibitemShut {NoStop}%
\bibitem [{\citenamefont {Loxley}\ and\ \citenamefont
  {Stamps}(2006)}]{loxley2006theory}%
  \BibitemOpen
  \bibfield  {author} {\bibinfo {author} {\bibfnamefont {P.~N.}\ \bibnamefont
  {Loxley}}\ and\ \bibinfo {author} {\bibfnamefont {R.~L.}\ \bibnamefont
  {Stamps}},\ }\href@noop {} {\bibfield  {journal} {\bibinfo  {journal}
  {Physical Review B}\ }\textbf {\bibinfo {volume} {73}},\ \bibinfo {pages}
  {024420} (\bibinfo {year} {2006})}\BibitemShut {NoStop}%
\bibitem [{\citenamefont {Winter}(1961)}]{winter1961bloch}%
  \BibitemOpen
  \bibfield  {author} {\bibinfo {author} {\bibfnamefont {J.~M.}\ \bibnamefont
  {Winter}},\ }\href@noop {} {\bibfield  {journal} {\bibinfo  {journal}
  {Physical Review}\ }\textbf {\bibinfo {volume} {124}},\ \bibinfo {pages}
  {452} (\bibinfo {year} {1961})}\BibitemShut {NoStop}%
\bibitem [{\citenamefont {Garcia-Sanchez}\ \emph {et~al.}(2015)\citenamefont
  {Garcia-Sanchez}, \citenamefont {Borys}, \citenamefont {Soucaille},
  \citenamefont {Adam}, \citenamefont {Stamps},\ and\ \citenamefont
  {Kim}}]{garcia2015narrow}%
  \BibitemOpen
  \bibfield  {author} {\bibinfo {author} {\bibfnamefont {F.}~\bibnamefont
  {Garcia-Sanchez}}, \bibinfo {author} {\bibfnamefont {P.}~\bibnamefont
  {Borys}}, \bibinfo {author} {\bibfnamefont {R.}~\bibnamefont {Soucaille}},
  \bibinfo {author} {\bibfnamefont {J.-P.}\ \bibnamefont {Adam}}, \bibinfo
  {author} {\bibfnamefont {R.~L.}\ \bibnamefont {Stamps}}, \ and\ \bibinfo
  {author} {\bibfnamefont {J.-V.}\ \bibnamefont {Kim}},\ }\href@noop {}
  {\bibfield  {journal} {\bibinfo  {journal} {Physical review letters}\
  }\textbf {\bibinfo {volume} {114}},\ \bibinfo {pages} {247206} (\bibinfo
  {year} {2015})}\BibitemShut {NoStop}%
\bibitem [{\citenamefont {Garcia-Sanchez}\ \emph {et~al.}(2014)\citenamefont
  {Garcia-Sanchez}, \citenamefont {Borys}, \citenamefont {Vansteenkiste},
  \citenamefont {Kim},\ and\ \citenamefont {Stamps}}]{GarciaSanchez:2014dw}%
  \BibitemOpen
  \bibfield  {author} {\bibinfo {author} {\bibfnamefont {F.}~\bibnamefont
  {Garcia-Sanchez}}, \bibinfo {author} {\bibfnamefont {P.}~\bibnamefont
  {Borys}}, \bibinfo {author} {\bibfnamefont {A.}~\bibnamefont
  {Vansteenkiste}}, \bibinfo {author} {\bibfnamefont {J.-V.}\ \bibnamefont
  {Kim}}, \ and\ \bibinfo {author} {\bibfnamefont {R.~L.}\ \bibnamefont
  {Stamps}},\ }\href@noop {} {\bibfield  {journal} {\bibinfo  {journal}
  {Physical Review B}\ }\textbf {\bibinfo {volume} {89}},\ \bibinfo {pages}
  {224408} (\bibinfo {year} {2014})}\BibitemShut {NoStop}%
\end{thebibliography}%
\end{document}